\documentclass[aps,twocolumn,pre,nofootinbib]{revtex4}
\usepackage{graphicx}
\usepackage{psfrag}
\usepackage{latexsym}
\usepackage{amsmath}
\usepackage{amssymb}
\usepackage{bm}

\newcommand{\bee}{\begin{eqnarray}}

\newcommand{\m}{\mbox{\boldmath $\mu$}}
\newcommand{\bF}{\mbox{\boldmath $F$}}
\newcommand{\nnabla}{\mbox{\boldmath $\nabla$}}
\newcommand{\eee}{\end{eqnarray}}
\newcommand{\ba}{\begin{array}}
\newcommand{\ea}{\end{array}}
\newcommand{\bc}{\begin{center}}
\newcommand{\ec}{\end{center}}
\newcommand{\bi}{\begin{itemize}}
\newcommand{\ei}{\end{itemize}}
\newcommand{\br}{\mbox{\boldmath $r$}}

\begin{document}

\title{A class of periodic and quasi-periodic trajectories of particles settling under gravity in a viscous fluid}
\author{Maria L. Ekiel-Je\.zewska}
\affiliation{Institute of Fundamental Technological Research, Polish Academy of Sciences, Pawi\'{n}skiego 5B, 02-106 Warsaw, Poland}
\date{\today}
\begin{abstract}
We investigate regular configurations of a small number 
of non-Brownian particles settling under gravity in a 
viscous fluid. The particles do not touch 
each other and can move relative to each other.  
The dynamics is analyzed in the point-particle 
approximation. A family of regular configurations is found 
with periodic oscillations of all the settling 
particles. The oscillations are shown to be robust 
under some out-of-phase rearrangements of the particles.
In the presence of an additional 
particle above such a regular configuration, the particle periodic trajectories are 
horizontally repelled from the symmetry axis, 
and flattened vertically. The results are used 
to propose a mechanism how a spherical cloud, made of a large 
number of particles distributed at random, evolves 
and destabilizes. 
\end{abstract}
\maketitle

\section{Introduction}
Periodic motions of a small number of particles attract a lot of interest because of their fundamental significance and their importance to understand the Stokesian dynamics of many-particle systems at random configurations \cite{Russel,Janosi,Stark,withUbbo1,withUbbo2}. For non-Brownian particles settling gravitationally in a viscous fluid under low-Reynolds-number, several classes of regular configurations oscillating periodically have been found and analyzed \cite{Hocking,Caflisch,Tory1,Tory2,Golubitsky,Lim,Snook,Mullin,Shelley,Gubiec,Lifetime,Adelaide}. 

It turns out that periodic trajectories can be essential for the dynamics of particles at random configurations.
In Ref. \onlinecite{Janosi}, the dynamics of three point-particles, initially at a random configuration, has been analyzed, and a chaotic scattering has been found.  
It has been shown that three close particles (both point-like \cite{Janosi} and spherical \cite{Lifetime}) circulate together 
before destabilizing into a faster pair and a slower singlet, and the interaction time is very sensitive to the initial conditions. In Ref. \onlinecite{Janosi}, the observed chaotic scattering of point-particles has been associated (without a proof) with the existence of an unknown unstable periodic relative trajectory.  For three spherical particles, 
such periodic trajectories indeed have been found \cite{Adelaide}. For random initial configurations of three particles, the shape of the relative trajectories has been shown to resemble the shape of the periodic ones \cite{Adelaide}. 
A group of three particles at a random configuration destabilizes 
when the system is sufficiently separated from such a  
periodic orbit. 

The question arises if a similar mechanism - the existence of a certain periodic relative trajectory  - can be applied to progress in understanding the dynamics of suspension drops sedimenting in a viscous fluid, i.e. swarms of particles randomly distributed in a spherical volume of the same fluid. 
The particles inside a sedimenting suspension drop
circulate and stay together for a long time,
 with the average particle and fluid velocity fields the same as in case of a more dense fluid drop \cite{Hadamard,Rybczynski,NitscheBatchelor,Ekiel}.
An initially spherical suspension drop later on 
slowly flattens and 
expands, forming a torus and occasionally leaving single 
particles behind (which, as slower, form  a thin long `tail' above the drop). Then, 
suddenly, the drop breaks into two (or sometimes more) 
fragments which repeat the same evolution pattern. The lifetime of a suspension drop is very sensitive to the initial conditions;  it can vary by orders of magnitude \cite{Mylyk2}. 

Dynamics of sedimenting swarms of particles %
has been extensively investigated 
experimentally and numerically, with the use of different 
methods~\cite{Mylyk2,NitscheBatchelor,Adachi,MeileSchaflinger,Abade,Bosse,Metzger,Koch,Alabrudzinski,Mylyk1}. 
However, it is still not clear what is the reason for the observed evolution pattern.
Are particles left behind 
the drop (as `a tail') if sufficiently separated from a
periodic orbit? Can the escaping particles cause the change of shape of the drop as the result of the hydrodynamic interaction? 
Does the wide range of the observed suspension-drop lifetimes result from a similar mechanism as the wide range of the three-particle cluster lifetimes? To address these open questions, the first step is to find and analyze families of periodic or quasi-periodic motions of such a number of particles, which can vary from a small to a very large value.

Therefore, the goal of this work is to construct %
such regular arrays of point-particles (in a geometry which resembles the shape of a sedimenting suspension drop), which oscillate while falling downward, and then destabilize,  
and to analyze 
basic properties of their periodic and quasi-periodic trajectories. In addition to the fundamental aspects of the results, such simple models can help to understand basic features of the %
 sedimenting suspension-drop dynamics, related to the above questions. 

\begin{figure*}
\hspace{-0.3cm} 
\parbox[t]{4.2cm}{\vspace{1.2cm}
\includegraphics[width=4.2cm]{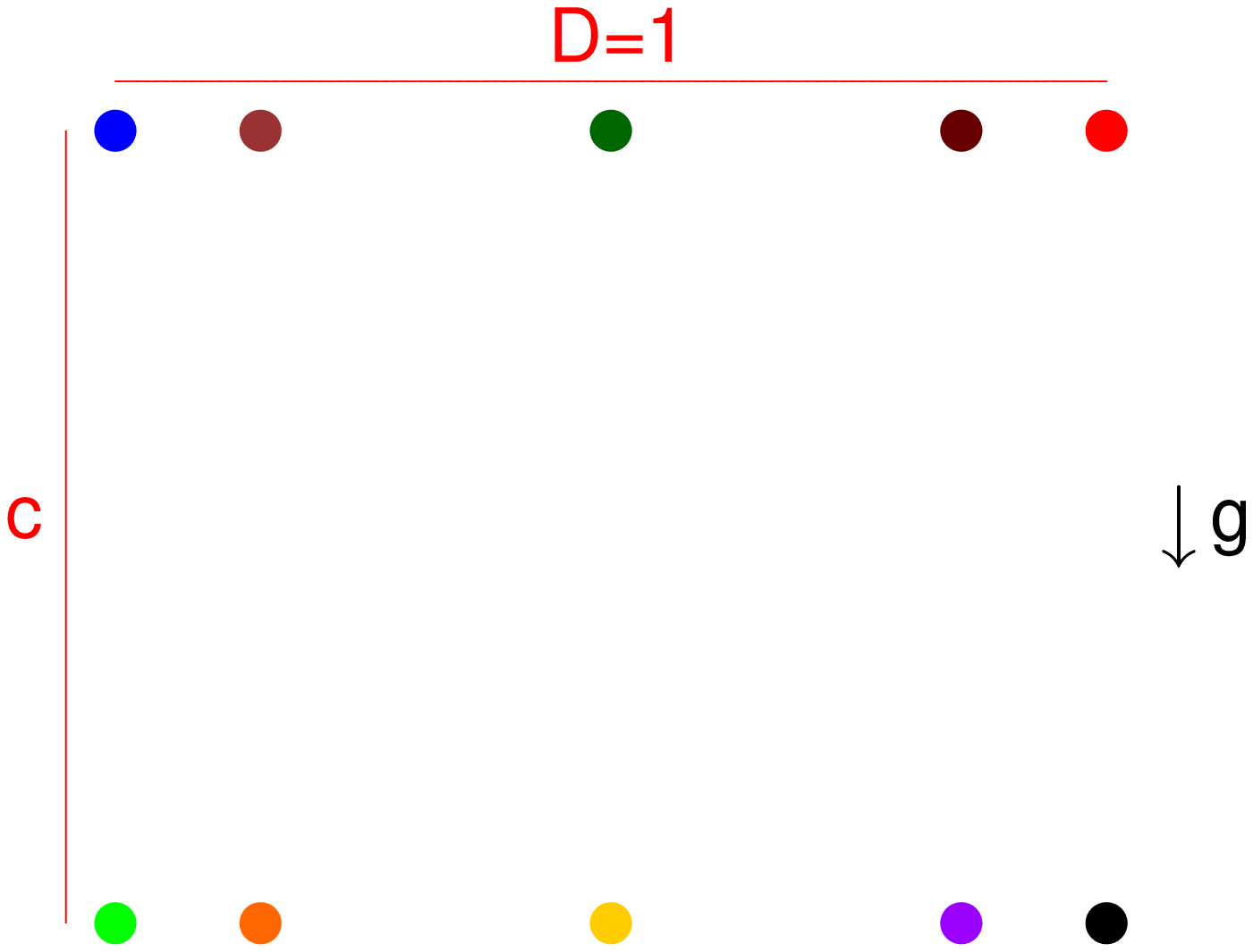}}\hspace{0.7cm}
\parbox[t]{6cm}{\vspace{0pt}\includegraphics[width=6cm]{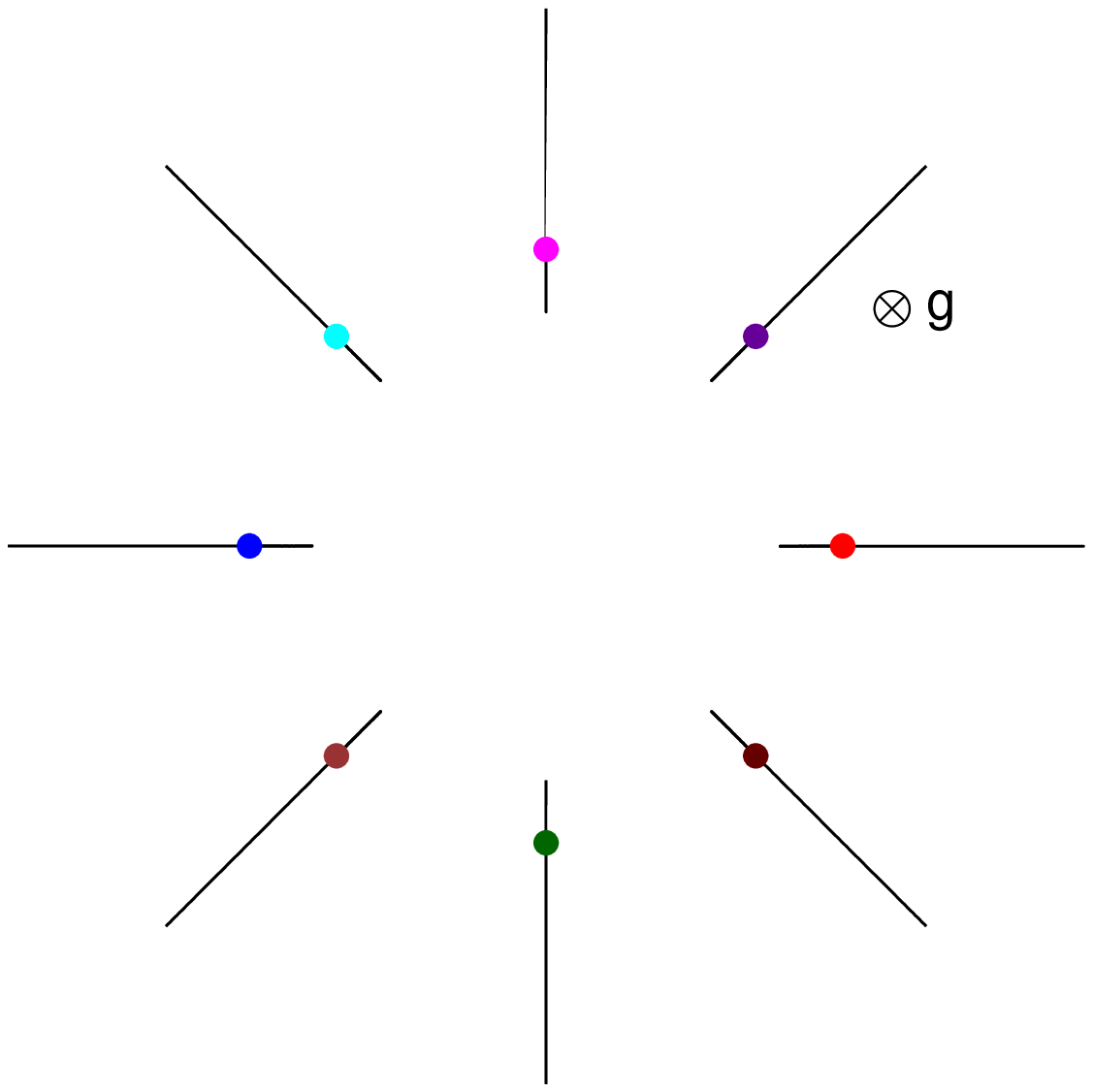}}\hspace{0.6cm}
\parbox[t]{6cm}{\vspace{.8cm}
\includegraphics[width=6cm]{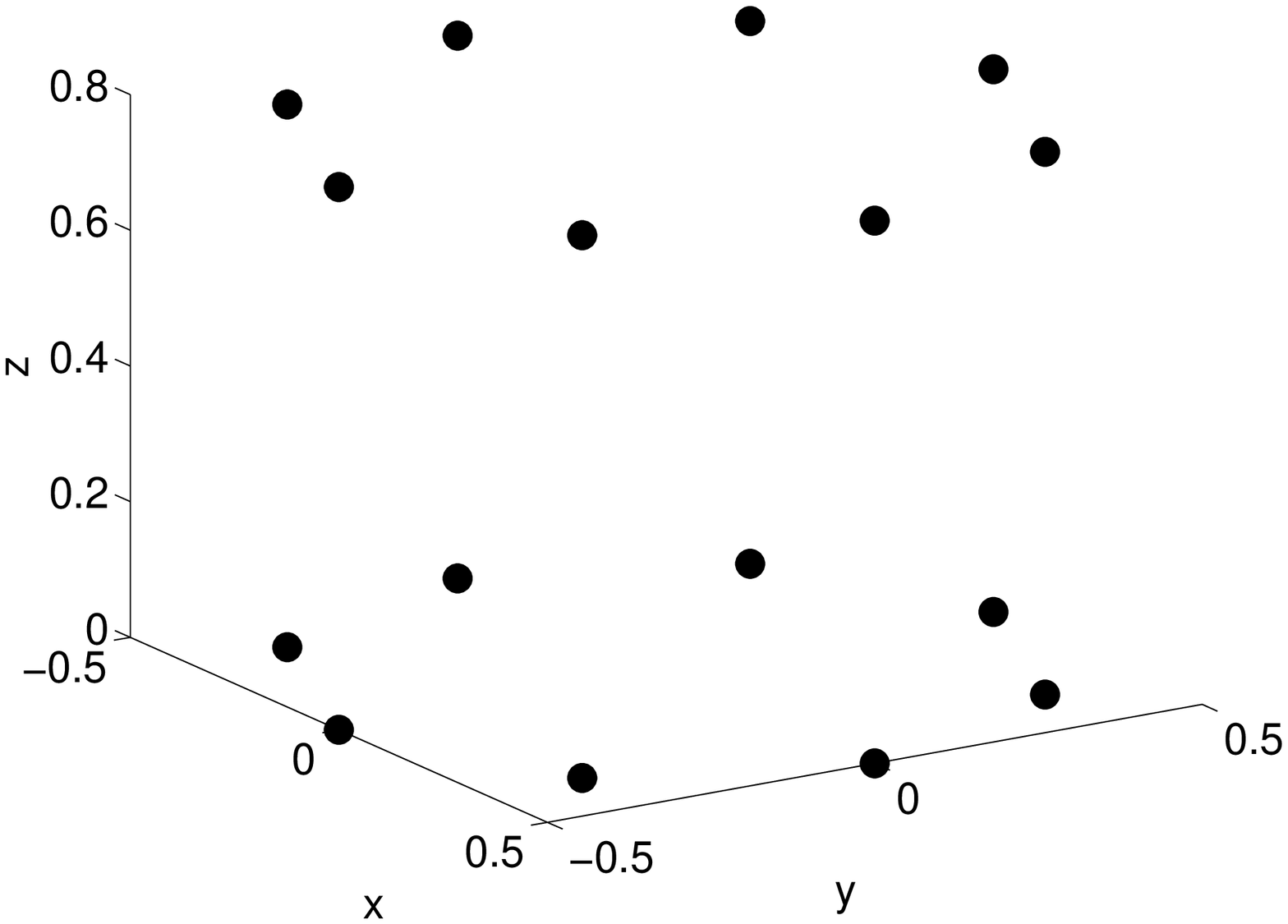}} 
\caption{Initial positions of 16 particles (dots) for c=0.8: side,
top and 3-D views. Solid lines at the middle panel: trajectories.
}\label{in1}
\end{figure*}

There is a lot of biological, medical, geological and industrial contexts were Stokesian dynamics of sedimenting clouds of particles is relevant; for example, colonies of bacteria or algae (including their periodic motions \cite{Goldstein}), clusters of particles in wastewater \cite{waste}, powders or sprays in human lungs \cite{med}. For groups of small non-Brownian particles sedimenting in water-based systems, the Reynolds number is typically much smaller than unity. E.g., for a group of one hundred particles, with the relative particle-fluid density equal to two, and particle radii equal to 5 $\mu$m, the Reynolds number Re$\approx 10^{-2}$.

The outline of the paper is the following. Sec.~\ref{II} contains presentation of the theoretical framework (the point-particle model). In Sec.~\ref{III}, the dynamics of regular configurations of particles is evaluated.  
A new family of periodic relative trajectories of the particles is analyzed. 
In Sec.~\ref{Pert}, these oscillations are shown to be robust under some out-of-phase rearrangements of the initial positions. Sec.~\ref{IV} illustrates how periodic motion of the particles, which form a regular configuration, is influenced by the presence of an additional particle above. Conclusions are presented in Sec.~\ref{V}. In Appendix~\ref{AA}, approximate dynamics of flat regular configurations of $2N$ particles is constructed and solved analytically. In Appendix~\ref{A}, the point-particle oscillations are shown to well-reproduce the periodic motion of spheres. 

\section{Theoretical framework}\label{II}
Assume that $K$ point-particles, located at $\br_i(t)$, with $i=1,...,K$, move in a fluid of viscosity $\eta$ under identical 
gravitational 
forces $\bF$. The fluid velocity ${\bf v}$ and pressure 
$p$ satisfy the Stokes equations,
\bee \eta {\bf \nnabla}^2 
{\bf v}(\br) -{\bf \nnabla} p(\br) &=& -\bF \sum_{i=1}^{K}\delta(\br-\br_i), \\
{\bf \nnabla} \cdot {\bf v}(\br) &=& 0.
\eee
In the reference frame moving with a single particle, the particle positions ${\br}_i(t)$, $i=1,...,K$,  satisfy the following evolution equations,
\bee
{\dot{\br}}_i(t)  &=& \left[ \sum_{k\ne i}^K\m_{ik}\right] \cdot  \bF, 
\hspace{1cm} i=1,...,K,\label{dy}
\eee
where 
the mobility 
$\m_{ik}$ 
is given by the Oseen tensor \cite{Kim-Karrila},
\bee
\m_{ik}&=&\frac{1}{8\pi\eta r_{ik}} 
({\bf I} + \hat{\br }_{ik} \hat{\br }_{ik})\hspace{0.4cm}\mbox{for }i\ne k,
\eee
with $\hat{\br }_{ik}=(\br_i-\br_k)/r_{ik} 
$ and $r_{ik}=|\br_i-\br_k|$. 
The frame of reference is chosen in such a way that the $z$-axis is anti-parallel to gravity, i.e. ${\bF}/|{\bF}|=(0,0,-1)$. 

Eqs~\eqref{dy} are solved numerically by the Adams-Bashforth-Moulton integration method (the {\it ode113} solver in {\sc matlab}). 

\section{Family of periodic solutions}\label{III}

Hocking analyzed oscillations of four point-particles settling under gravity in a vertical plane~\cite{Hocking}. 
His initial configurations %
can be modified to start from point particles located at vertices of 
a rectangle with vertical and horizontal sides, all located in the same vertical plane. 

In this work, a generalized initial configuration is considered:  
 $2N$ point-particles located at vertices of 
a regular right prism, which consists of $`$twin' horizontal regular $N$-polygons, with each particle exactly above or below another one, separated by a distance $c$,
\bee
{\br }_{k}(0)\!\!\!&=& \!\!\!\left\{ \ba{l}\!\!\!(\frac{1}{2}\!\cos \!\frac{2\pi (k\!-\!1)}{N}, \frac{1}{2}\!\sin \!\frac{2\pi (k\!-\!1)}{N}, 0)\hspace{0.1cm} \mbox{for }k\!=\!1,...,N,\\\\
\!\!\!(\frac{1}{2}\!\cos \!\frac{2\pi (k\!-\!1)}{N}, \frac{1}{2}\!\sin \!\frac{2\pi (k\!-\!1)}{N}, c)\hspace{0.1cm} \mbox{for }k\!=\!\!N\!\!+\!\!1,...,2N.\!\!\ea\right.\nonumber\\\label{in}
\eee

The frame of reference is chosen in such a way that 
the $xz$-plane contains the initial positions of two or four particles, for $N$ odd or even, respectively, and $z$ is along the rotational symmetry axis. 
The length unit $D$ is twice the initial distance of a particle from the symmetry axis, and
the time unit is $8 \pi \eta D^2/F$. 

Owing to the symmetry  with respect to rotations by $2\pi/N$ of the initial
configurations specified in Eq.~\eqref{in}, the periodic relative motion of the particles takes place in vertical planes which include the center-of-mass of the whole group (in particular, in the $xz$ plane), and the shape of all the relative trajectories is identical.
In the reference frame of the center-of-mass, a particle and its twin %
follow the same periodic trajectory. 

\begin{figure}[h]
\psfrag{i}{}
\psfrag{x}{$x_{i}$}
\psfrag{z}{$z_{i}$}
\hspace{-0.8cm} \includegraphics[width=7.5cm]{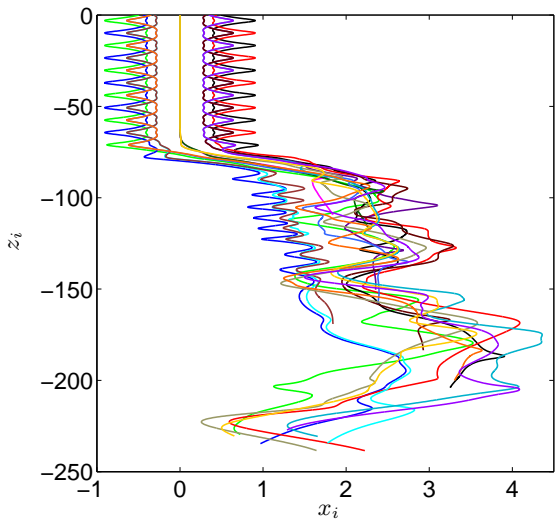}
\vspace{-0.2cm}
\caption{Trajectories of 16 particles initially located as shown in Fig.~\ref{in1}, with $c=0.8$ (side view).} \label{tra}
\end{figure}

As an example, we consider evolution of the initial configuration of $2N\!=\!16$ particles, shown in Fig.~\ref{in1}, with $c\!=\!0.8$. 
All the particle trajectories, observed during $t=25$ in the laboratory frame of reference (which moves with a single-particle velocity), are plotted in Fig.~\ref{tra}. 

There appear two characteristic stages of the evolution. In the first one, for $t\lesssim 3$ (five periods), 
 the particles 
perform periodic oscillations while falling downward. Then, the cluster destabilizes, and the motion is not regular any more. 
One by one, the particles are lost behind the cluster. This process is slow: at $t=25$, still a half of the particles stay relatively close to each other. 

In the computations, the relative $\Delta_r$ and absolute $\Delta_a$ error tolerances in the numerical integration routine were equal to $10^{-12}$. It has been checked that the destabilization time, period of the oscillations and particle periodic trajectories are independent of error tolerances, providing that they are small enough, i.e. $\Delta_r \le 10^{-3}$ and $\Delta_a\le 10^{-6}$. On the contrary,  the destabilization pattern is very sensitive to tiny perturbations, even as small as a small change of the error tolerance in the numerical solvers. The pattern shown in Fig.~\ref{tra}, with the
center of mass moving towards positive y values, 
is just an example of many possible ways of the system break-up, observed numerically for different parameters of the numerical procedure, or small perturbations (including e.g. the center of mass moving towards positive or zero y values).  

\begin{figure}[h]
\psfrag{CM}{}
\psfrag{y}{$x_{i,CM}$}
\psfrag{z}{$z_{i,CM}$}
\hspace{-0.8cm}\includegraphics[width=8cm]{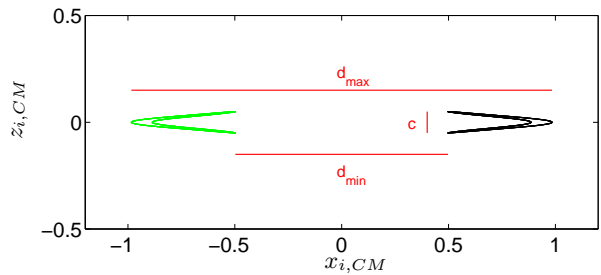}
\vspace{-0.5cm}
\caption{Trajectories in the center-of-mass frame for
$c\!=\!0.1$.} 
\label{notation}
\end{figure}

We now adopt the center-of-mass frame and a vertical plane in which the motion of four particles takes place. We investigate how shapes of periodic trajectories  depend on $c \le 3.5$.
The characteristic parameters of the group trajectories 
are their height $c$, and the group maximal and minimal width, $d_{\text{max}}$ and $d_{\text{min}}$, respectively (twice the maximal and minimal excursion from the symmetry axis),  %
as shown in Fig.~\ref{notation}. We also evaluate the aspect-ratio of the group trajectories,
\bee
p&=&c/d_{\text{max}}.
\eee

\begin{figure}[h]
\psfrag{CM}{}
\psfrag{y}{$x_{i,CM}$}
\psfrag{z}{$z_{i,CM}$}
\hspace{-0.8cm}\includegraphics[width=8cm]{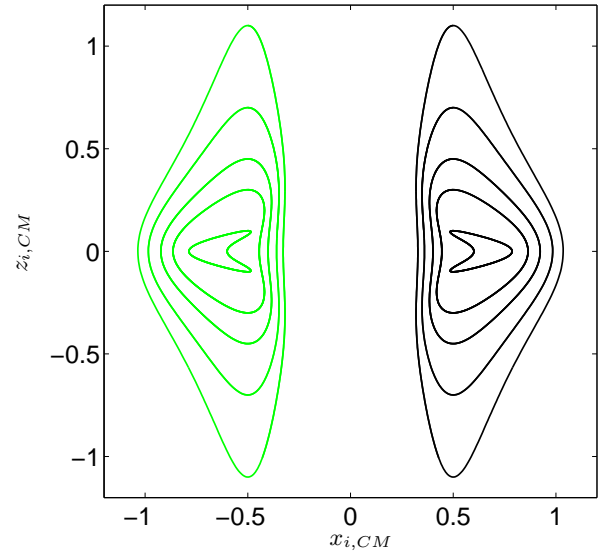}
\vspace{-0.3cm}
\caption{Trajectories in the center-of-mass frame, for 
$c=2.2, 1.4, 0.9, 0.6, 0.2$. The smaller $c$, the shorter the trajectory and the smaller its width, 
$(d_{\text{max}}\!-\!d_{\text{min}})/2$.}\label{trCMa}
\end{figure}

\begin{figure}[!h]
\vspace{-0.4cm}
 \includegraphics[width=8.2cm]{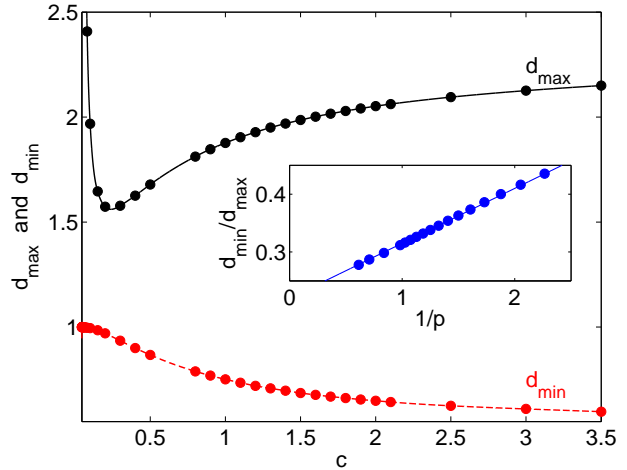}
\vspace{-0.4cm}
\caption{The maximal and minimal width of the group trajectories, $d_{\text{max}}$ and $d_{\text{min}}$, versus $c$, for $c\ge 0.04481$.  
Inset: for $c \ge 0.8$, the ratio $d_{\text{min}}/d_{\text{max}}$ (symbols) scales as $0.095/p+0.22$ (solid line).}\label{wandh}\vspace{-0.5cm}
\end{figure}

\begin{figure*}
\psfrag{CM}{}
\psfrag{y}{$x_{i,CM}$}
\psfrag{z}{$z_{i,CM}$}
\includegraphics[width=15.6cm]{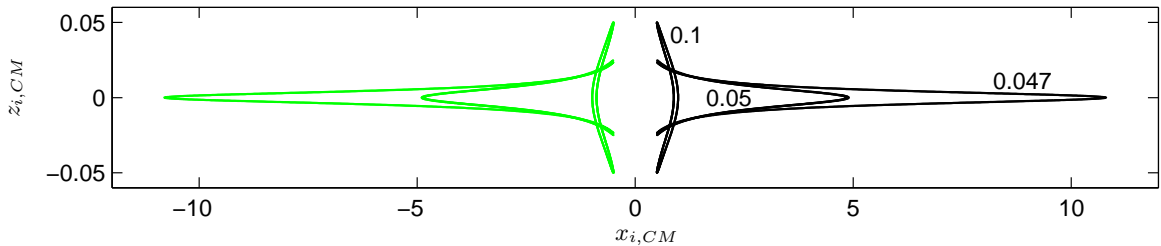}\\
\vspace{-0.3cm}
\caption{Trajectories in the center-of-mass frame, for small values of $c= 0.1, 0.05, 0.047$. 
The smaller $c$, the longer and wider the trajectory. Note a different scale on each axis.}\label{trCM}
\end{figure*}

In Figs~\ref{trCMa} and \ref{wandh},
we illustrate that when $c$ decreases  
down to $c \approx $0.2,  
the width $d_{\text{max}}$ and the aspect ratio $p$
of group trajectories are getting smaller. The total arc length $L$ of a
closed trajectory 
decreases, and a successive trajectory is located inside the previous one. When $c$ still decreases, 
$d_{\text{min}}$, 
``the diameter of the hole'', increases up to one (the upper limit determined by the initial conditions) and 
the width $d_{\text{max}}$ of the group trajectories increases
rapidly,
see Figs~\ref{wandh} and~\ref{trCM}.

\begin{figure}[h]
 \includegraphics[width=8.2cm]{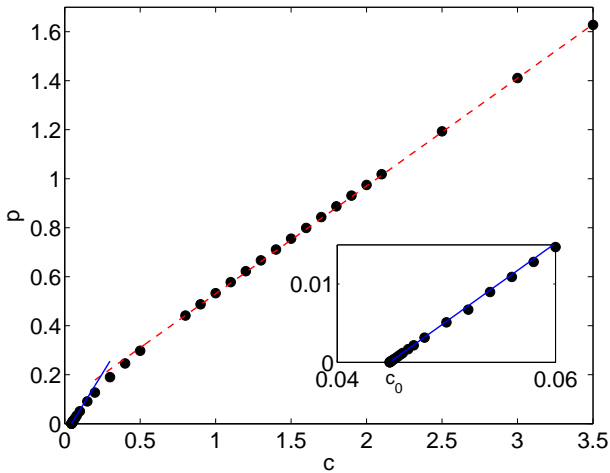}\vspace{-0.4cm}
\caption{The aspect ratio $p$ of the group trajectories as a function of $c$ (symbols). Straight lines: 
dashed (red online), $p\!=\!0.44c\!+\!0.09$,
and 
solid (blue online), $p=c\!-\!c_0$. 
}\label{ga}\vspace{-0.3cm}
\end{figure}

Different behavior for small at large values of $c$ is also seen in Fig.~\ref{ga}.
In the whole range of $c$, the aspect ratio $p$ of the group trajectories 
decreases when $c$ is decreased, but for $c\gtrsim 0.3$, the slope is less steep than for smaller values of $c$.
\begin{figure}[!h]
 \includegraphics[width=8.5cm]{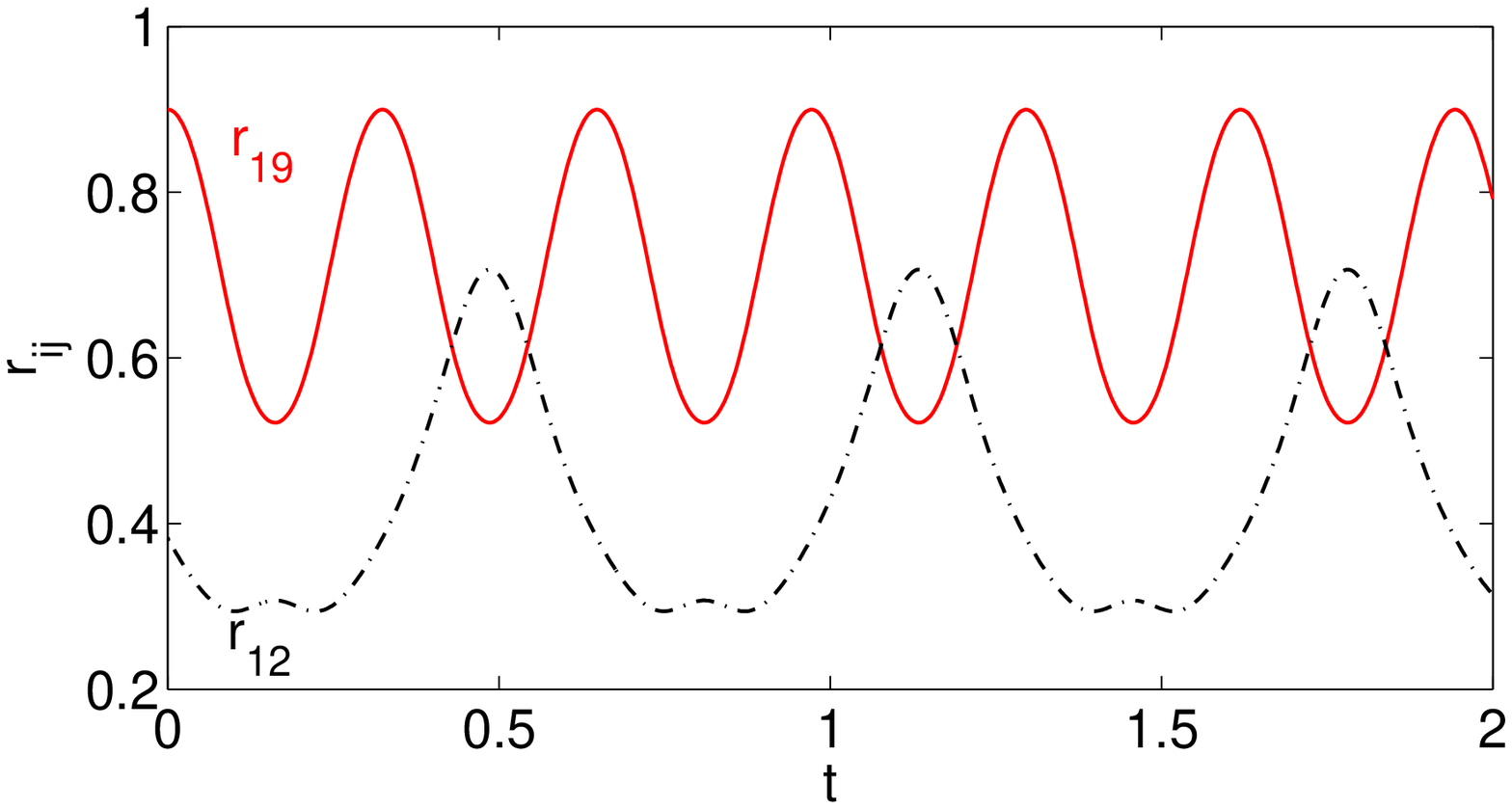}
\\ \vspace{-0.1cm}
 \includegraphics[width=8.5cm]{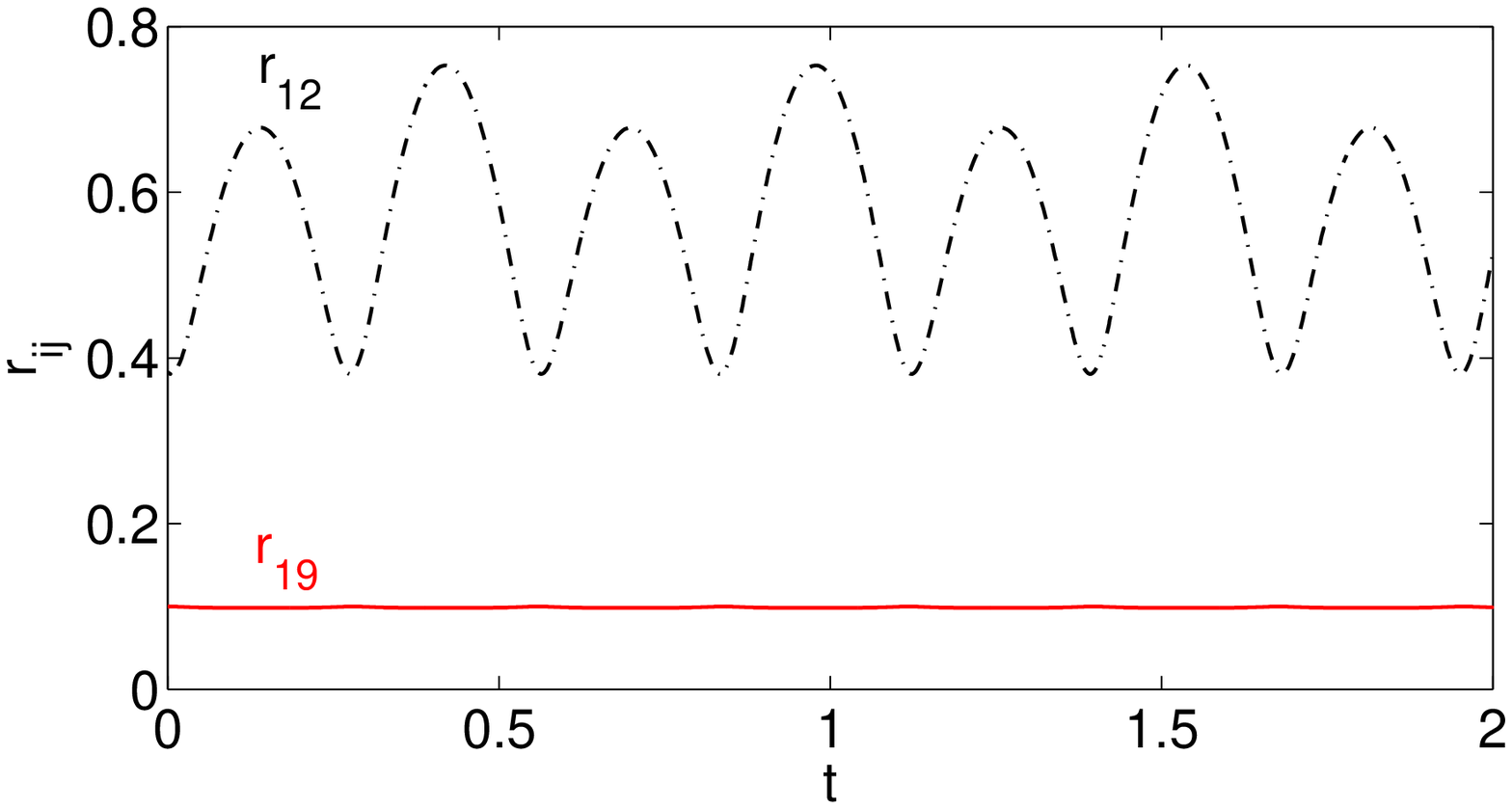}
\vspace{-0.7cm}
\caption{The time-dependent interparticle distance between the twin particles, $r_{19}$, and between the closest particles from the polygon, $r_{12}$. Top: $c=0.9$. Bottom: $c=0.1$.}\label{dist}
\end{figure}

The qualitative change of the dynamics takes place for such initial configurations, for which the distance $c$ between a particle and its twin just above or below it (e.g. the particles 1 and 9), is of the same order as the distance $\sin(\pi/8)\approx 0.38$ between the closest particles from the same horizontal polygon (e.g. the particles 1 and 2). 

To explain the reason of the dynamics change, in Fig.~\ref{dist} 
we compare the time-dependent interparticle distance $r_{19}$ between the twin particles to the distance $r_{12}$ between the closest neighbors from the same polygon (with $r_{ij}=|\br_i-\br_j|$), for a small and a large value of $c$. For larger values of $c$, e.g. $c=0.9$, 
at most of the times, $r_{19}>r_{12}$, but it also happens that $r_{19}<r_{12}$. On the time-average, the hydrodynamic interactions between the closest particles from the same polygon are stronger than between the twin particles.

For smaller values of $c$, e.g. $c=0.1$, the twin particles are always much closer to each other than to any other particle, and therefore, they interact with each other much stronger; in a sense, they are hydrodynamically ``teamed-up''.

\begin{figure}[h]
 \includegraphics[width=8.2cm]{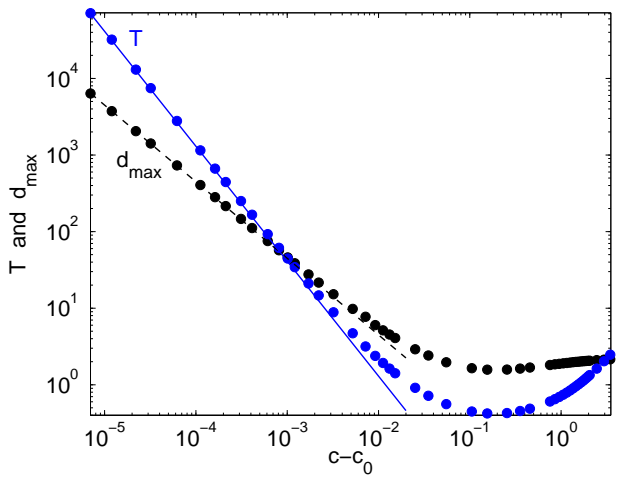}\vspace{-0.2cm}
\caption{The period $T$ and the cluster width $d_{\text{max}}$ versus $c\!-\!c_0$ (symbols). Straight lines: $T\!=\!0.0013/(c\!-\!c_0)^{1.5}$ (solid) and $d_{\text{max}}\!=\!c_0/(c\!-\!c_0)$ (dashed).}\label{loglog}
\end{figure}

For so small values of $c$, we observe in Fig.~\ref{trCM} that the arc length $L$ and width $d_{\text{max}}$ of a closed center-of-mass trajectory, and also period $T$ of the oscillations 
increase significantly when $c$ decreases even a little. The pairs of ``teamed-up'' twin particles tend to escape, but eventually are stopped by interactions with the other particles. Is there a critical value $c=c_0$ of the aspect ratio, where they all become infinite? To check, we assume a power law divergence, 
\bee
T \!\sim \!A/(c-c_0)^{\alpha} \mbox{ and } d_{\text{max}} \!\sim\! B/ (c-c_0)^{\beta}, \hspace{0.3cm}\mbox{for } c\rightarrow c_0,\nonumber \\\label{Tc-c0}
\eee
and search for $c_0, 
\alpha, \beta$ by plotting  $T$ and $d_{\text{max}}$ versus $(c-c_0)$ in the log-log scale, see Fig.~\ref{loglog}.

In this way, we find the critical exponents,
\bee
\alpha \approx 1.50,  \hspace{0.5cm}\beta \approx 1.00,\label{ww}
\eee
and the critical value of the aspect ratio~$c$,
\bee
c_0 &=& 0.044788....\label{c0_16}
\eee

For all the investigated values of $2N$ (including the benchmark solution for $2N=4$), the relations \eqref{Tc-c0} are also valid, with the same $\alpha$ and $\beta$, but different values of $c_0$. In Appendix \ref{A}, the power law scalings \eqref{Tc-c0}-\eqref{ww} and values of the critical aspect ratios $c_0$ are derived from approximate dynamics of the regular prisms with $2N$ particles.

\begin{figure}[h]
\psfrag{y}{$x_i$}
\psfrag{z}{$z_i$}
\vspace{0.2cm} \includegraphics[width=7.5cm]{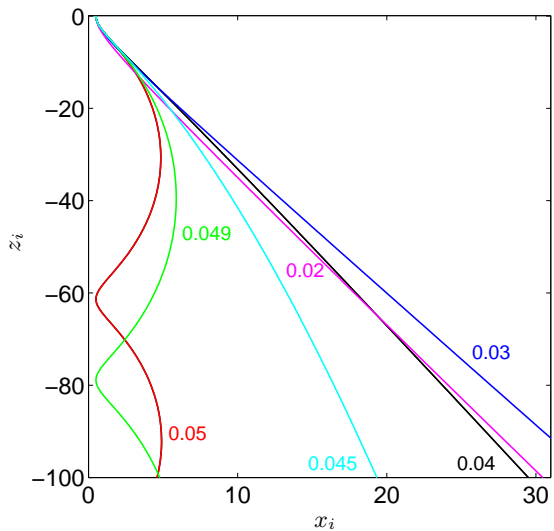}\vspace{-0.2cm}
\caption{Two types of the particle trajectories: periodic oscillations ($c=0.05, 0.049, 0.045$) and separation into pairs without oscillations ($c=0.04, 0.03, 0.02$).}\label{per-sep}
\end{figure}

For $c<c_0$, there is no periodic motions; the group from the beginning splits into $N$ pairs of particles. The comparison of the trajectories without and with periodic oscillations is performed in Fig.~\ref{per-sep}. For $c<c_0$, the slope of a trajectory monotonically decreases to a constant value, which is smaller than the minimal slope reached at periodic trajectories.

\section{Perturbations}\label{Pert}
In this section, we perturb the initial particle configurations given by Eq.~\eqref{in} 
and investigate what is the resulting change of the dynamics. We analyze two examples of perturbations.

First, we rotate the upper polygon by $\pi/N$, leaving it in its original plane; the perturbed initial positions are,
\bee
{\br }_{k}(0)\!\!\!&=& \!\!\!\left\{ \ba{l}\!\!\!(\frac{1}{2}\!\cos \!\frac{2\pi (k-1)}{N}, \frac{1}{2}\!\sin \!\frac{2\pi (k-1)}{N}, 0)\hspace{0.1cm} \mbox{for }k\!=\!1,...,N,\\\\
\!\!\!(\frac{1}{2}\!\cos \!\frac{\pi (2k\!-\!1)}{N}, \frac{1}{2}\!\sin \!\frac{\pi (2k\!-\!1)}{N}, c)\hspace{0.1cm} \mbox{for }k\!=\!N\!\!+\!\!1,...,2N.\!\!\!\ea\right.\nonumber\\\label{in_prz}
\eee

The number of trajectories and the number of vertical planes of the motion increase by a factor of two in comparison to the unperturbed case. In the reference frame moving with the center-of-mass of the system, 
each particle moves along its own trajectory, in contrast to the solutions presented in Sec.~\ref{III}, for which a pair of particles moves along the same trajectory.

\begin{figure}[h]
\psfrag{CM}{}
\psfrag{y}{$x_{i,CM}$}
\psfrag{z}{$z_{i,CM}$}
\vspace{0.2cm} \includegraphics[width=7.5cm]{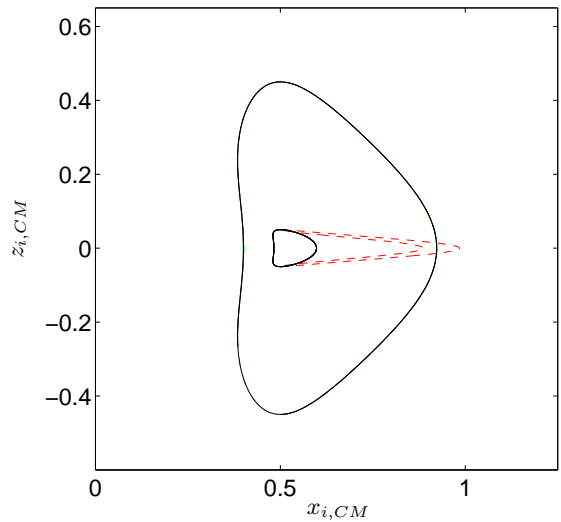}\vspace{-0.2cm}
    \caption{In the center-of-mass frame, the particle trajectories with the perturbed initial positions, 
Eq. \eqref{in_prz}, (solid lines), and with the unperturbed ones, Eq.~\eqref{in}, (dashed lines), are superimposed for $c\!\!=\!\!0.9$ and differ from each other significantly for $c\!\!=\!\!0.1$.%
}\label{per1}
\end{figure}

The shape of the particle trajectory in the center-of-mass frame is the practically same as for the unperturbed solution if $c$ is large, and significantly different when $c$ is small, as illustrated in Fig.~\ref{per1} for $2N\!\!=\!\!16$. For small values of $c$,  the perturbed
 initial condition does not lead to such a strong hydrodynamic coupling of the twin particles as for the unperturbed solution, and as the result, the group trajectories are less wide.

The second type of the perturbation is applied to the initial configuration \eqref{in} with even N. We use values of $d_{\text{max}}$ and $d_{\text{min}}$ evaluated in Sec.~\ref{III} to 
construct a new initial condition, by shifting every second particle in the polygon to a position where it would be expected after one fourth of the period of the unperturbed solution. In this way, we want to test if there exist ``out-of-phase'' periodic oscillations. 
Explicitly, the initial conditions are,
\bee
{\br }_{k}(0)=\hspace{4cm}\nonumber\\
 \!\!\!\left\{ \ba{l}\!\!\!(\frac{1}{2}\cos \frac{2\pi (k\!-\!1)}{N}, \frac{1}{2}\sin \frac{2\pi (k\!-\!1)}{N}, 0)\hspace{0.1cm} \mbox{for }k\!=\!2,4,...,N,\\\nonumber\\
\!\!\!(\frac{1}{2}\!\cos \!\frac{2\pi (k\!-\!1)}{N}, \frac{1}{2}\!\sin \!\frac{2\pi (k\!-\!1)}{N}, c)\hspace{0.1cm} \mbox{for }k\!=\!N\!\!+\!\!2,N\!\!+\!\!4,...,2N,\!\!\!\\\nonumber\\
\!\!\!(\frac{d_{\text{max}}}{2}\!\cos \!\frac{2\pi (k\!-\!1)}{N}, \frac{d_{\text{max}}}{2}\!\sin \!\frac{2\pi (k\!-\!1)}{N}, \frac{c}{2})\hspace{0.1cm} \mbox{for }k\!=\!1,3,...,N\!\!-\!\!1,\\\\
\!\!\!(\frac{d_{\text{min}}}{2}\!\cos \!\frac{2\pi (k\!-\!1)}{N}, \frac{d_{\text{min}}}{2}\!\sin \!\frac{2\pi (k\!-\!1)}{N}, \frac{c}{2})\hspace{0.1cm} \mbox{for }k\!=\!N\!\!+\!\!1,...,2N\!\!-\!\!1,\!\!\!\!
\ea\!\!\!\right.\!\!\!\!\nonumber\\\!\!\!\!\!\!\label{ioop}
\eee
with even N.

For example, we show the results for the initial positions of $2N\!=\!16$ particles, and in Eq.~\eqref{ioop} we use $d_{\text{max}}\!=\!1.8468$ and $d_{\text{min}}\!=\!0.8028$ for $c\!=\!0.9$ and 
$d_{\text{max}}\!=\!1.9686$ and $d_{\text{min}}\!=\!1.7720$ for $c\!=\!0.1$. 

\begin{figure}[h]
\vspace{0.2cm} \includegraphics[width=8.6cm]{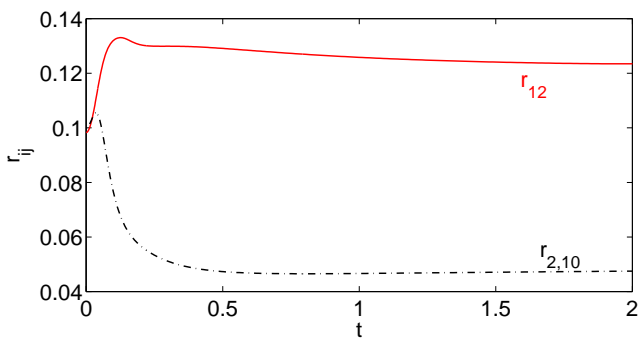}\vspace{-0.2cm}
\caption{The time-dependent distance 
 between the slower twin particles, $r_{19}$,
(solid line) and between the faster twin particles, $r_{2,10}$,
(dashed line), 
for the initial positions given by Eq.~\eqref{ioop} with $2N\!=\!16$ and $c\!=\!0.1$.}\label{oop1}
\end{figure}

For smaller values of $c$, e.g. 
$c\!=\!0.1$, the configuration from the very beginning 
separates into two groups: slower particles with odd indices (those which have been shifted)  and faster particles with even indices (unperturbed). The reason is that the distance between the those twin particles, which have been shifted, after a time becomes larger than between the unperturbed ones, as illustrated in Fig.~\ref{oop1}.

\begin{figure}[h]
\psfrag{CM}{}
\psfrag{y}{$x_{i,CM}$}
\psfrag{z}{$z_{i,CM}$}
\vspace{0.2cm} \includegraphics[width=7.5cm]{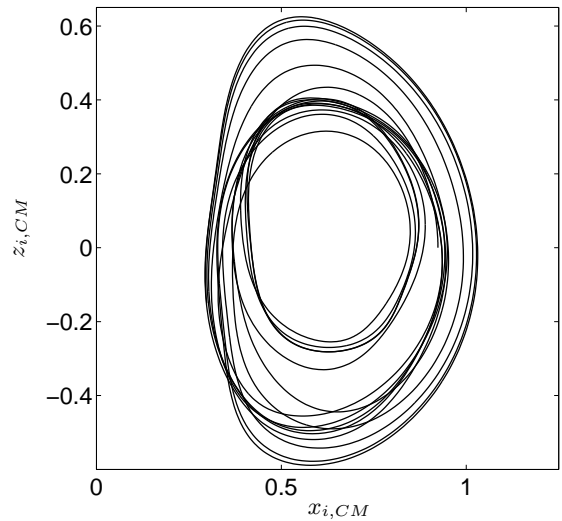}\vspace{-0.2cm}
\caption{The quasi-periodic particle trajectory in the center-of-mass frame, obtained for the initial positions given by Eq.~\eqref{ioop} with $2N\!=\!16$ and $c\!=\!0.9$, during time $0 \le t \le 10$.}\label{oop2}
\end{figure}

For 
larger values of $c$, the particles interact hydrodynamically with each other with a comparable strength, and form a single group for a long time (e.g. for $c=0.9$, almost 4 times longer than in the unperturbed case). The relative motion is quasi-periodic. For $c=0.9$, the particle trajectory in the center-of-mass frame is shown in Fig.~\ref{oop2}. Compare with the unperturbed trajectory in Fig.~\ref{per1} (the scale in both figures is the same).

Concluding, it has been shown that there exist perturbations of the regular configurations from Sec~\ref{III}, which lead to out-of-phase long-lasting quasi-periodic oscillations of all the particles. Such solutions are good candidates as seeds for a next, more complex and realistic generation of models of the dynamics of a random suspension drop. 

\section{A group of particles with ``a tail'' above}\label{IV}
We will now illustrate how periodic motions of particles in a regular configuration, investigated in Sec~\ref{III}, are modified by the presence of `a tail' made of an additional particle above.
Such a system is supposed to model a suspension drop and a tail of particles gradually lost from it. It is known that the particles, which will later separate out from a suspension drop, circulate along the exterior trajectories up to the top of the drop, become slower than the drop, and therefore are left behind the drop as `a tail' along the symmetry axis above the drop~\cite{NitscheBatchelor}.

In our model, the drop is represented by the regular configuration, which consists of 16 particles, and the tail from a singlet just above the center of mass of the group. The initial positions of 16 particles are given by Eq.~\eqref{in} with $c=0.9$, 
and shown in Fig.~\ref{in1}. 
 At $t=0$, and the 17th particle is placed at the symmetry axis of the regular group, at a small distance $z_0=1.25$ above its center-of-mass. 
The question is how the trajectories of the 16 particles, evaluated in their center-of-mass frame (and shown as the middle curves in Fig.~\ref{trCMa}), are modified by the presence of the 17$^{\mbox{th}}$ particle. Can a single particle significantly change periodic trajectories of all the 16 particles?

\begin{figure}[h]
\psfrag{CM,1-16}{}
\psfrag{y}{$x_{i,CM,1-16}$}
\psfrag{z}{$z_{i,CM,1-16}$}
\hspace{-0.8cm}\includegraphics[width=7.5cm]{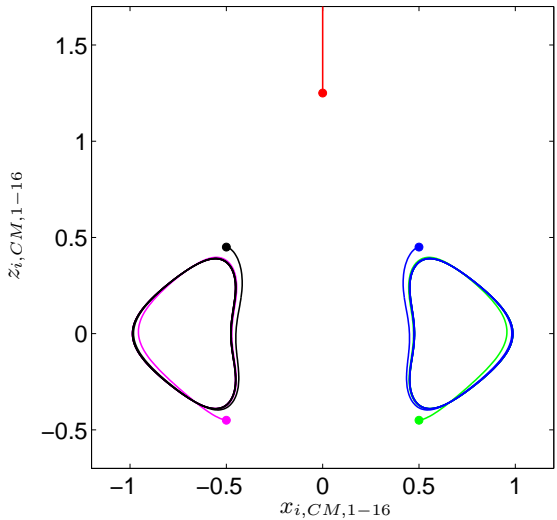}
\caption{Evolution of 17 particles, initially at the positions (indicated by dots)
described by 
Eq.~\eqref{in} with $c=0.9$ plus ${\bf r}_{17}=(0,0,1.7)$, in the center-of-mass frame of the regular configuration made of 16 particles. Trajectories of the other 12 particles look the same in their planes of the motion.
}\label{drop+tail}
\end{figure}
The trajectories in the center-of-mass frame of the regular configuration (CM,1-16) 
are shown in Fig.~\ref{drop+tail}. The trajectories of five particles are shown - all those which move in the $yz$-plane.
In the description of the axes, we indicated that the reference frame is moving with the center-of-mass of the 16 particles.  It is clear that with time, the shape of each closed trajectory changes.    It becomes less high, wider, and its distance from the symmetry axis increases. The same features are observed when the long-time part of the trajectory from Fig.~\ref{drop+tail} is compared to the corresponding trajectory in the absence of the 17$^{\mbox{th}}$ particle, see the curve with $c=0.9$ in Figs~\ref{trCMa} and \ref{per1}. 

\begin{figure}[h]
\includegraphics[width=3.8cm]{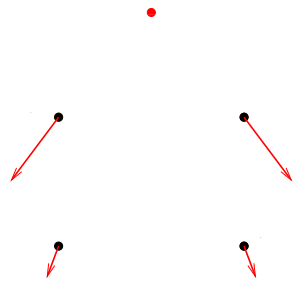}
\caption{Schematic explanation of the horizontal expansion and vertical flattening of the particle trajectories inside the group. Arrows (sketch): this part of the group-particle velocity, which comes from its interaction with the tail-particle.}\label{expl}
\end{figure}

The explanation is that the tail particle interacts hydrodynamically with the particles in the group above. 
The additional velocity of a particle from the group, gained owing to its interaction with the tail (the Oseen velocity generated by the single point-force), is schematically indicated by arrows in Fig.~\ref{expl}. As the result, 
the particles inside the group
are repelled horizontally from the group center, and 
attracted vertically to the central plane.  The closer the tail, the larger the effect.

\section{Conclusions}\label{V}
A new class of unstable periodic relative motions has been found: 
initially, $2N$ point-particles form a regular prism (two mirror 
horizontal regular polygons) with a vertical-to-horizontal 
aspect ratio $c$ larger than a critical value 
$c_0$. For $c<c_0$, the system from the beginning separates into pairs of the twin mirror particles. 
Such periodic motions have been observed 
for different values of $N$, with a smaller $c_0$ for a larger $N$.

For moderate value of $c$ %
and larger $N$, each particle is stronger coupled to the closest neighbors in the polygon than to its twin particle.
For smaller $c$ and smaller $N$, the inverse ordering of the coupling is observed, and in this case, approximate analytic solutions have been found. 

From the point of view of relevance to spherical systems of randomly distributed particles, the most interesting is the intermediate case, when all the particles interact with each other with a comparable strength, and the interparticle distances are of the same order of magnitude. Therefore, we focused on computing evolution of systems made of $2N\!=\!16, 32, 64$ particles.

By perturbing the regular configurations described above we obtained another family of periodic solutions, and a class of quasi-periodic, long-lasting, out of phase  oscillations of particles. 
We demonstrated 
that a single particle above the center of mass of the regular configuration repels particles horizontally from the center 
and attracts them vertically to the central horizontal plane of the configuration.

The results can be used as basic models to study a mechanism of deformation and destabilization of initially spherical suspension drops. The hypothesis to be checked is the following. The particles inside the drop `stay close' to a periodic or quasi-periodic trajectory. 
Owing to statistical fluctuations, particles which are too far from such an orbit, stop circulating and are left behind the drop as a tail above the center-of-mass of the drop. 
The tail interacts hydrodynamically with the particles inside the drop and causes the drop to flatten and expand horizontally, with a characteristic time scale $\tau_t$ of the drop 
deformation.
The larger time, the smaller value of $c$ for the corresponding model periodic or quasi-periodic 
solution of a regular configuration.
Destabilization time of a suspension drop can be associated with the 
 characteristic growth time  $\tau_d\!<< \!\tau_t$ of a certain periodic-orbit perturbation. 
 Such a mechanism is consistent with a wide range of the drop destabilization times, observed in experiments and simulations~\cite{Mylyk2}, just as it is in case of 3 particles only~\cite{Janosi,Lifetime}. 

\acknowledgments
This work was supported in part by the Polish National Science 
Centre grant 2011/01/B/ST3/05691. Scientific benefits from the 
activities of the COST Action MP1106 are acknowledged.

\appendix
\section{Power law scaling derived from approximate dynamics}\label{AA}
To justify the power law scalings \eqref{Tc-c0}-\eqref{ww}, we will now construct an approximate dynamics of the regular right prisms with $2N$ particles, assuming for simplicity that $N$ is even. 
The basic observation is that for small aspect ratios $c$, the distance between the twin particles (i.e. those which move along the same relative trajectory) is practically constant during the motion (see the plot of the distance $r_{19}$ in the bottom panel of Fig.~\ref{dist} and both curves in Fig.~\ref{oop1}). Therefore, the key point of the approximation is to recover 
the constant distance between each pair of the twin particles. 

Consider first a simpler example of $2N\!\!=\!\!4$ particles only. It will later become clear that the solution for this special case is generic for an arbitrary number of particles $2N$. 
In analogy to Ref.~\cite{Hocking}, we denote relative coordinates of the twin particles $1$ and 
$3$ as follows,
\bee
x=x_1+x_3,\hspace{0.5cm}y=x_1-x_3,\hspace{0.5cm}z=z_3-z_1,\label{no}
\eee
with the particle positions ${\bf r}_i=(x_i,0,z_i)$.

Starting from flat initial configurations given by Eq.~\eqref{in} with a small aspect ratio,
\bee
c<<1,
\eee 
and 
assuming that the twin particles stay closer to each other than to other particles, 
\bee
{y},{z} &<<& x,
\eee
we approximate Eqs~\eqref{dy} as 
\bee
\frac{dx}{dt} &=& 2\frac{yz}{c^3},\label{dxdt}\\
\frac{dy}{dt} &=& 2\frac{z}{x^2},\label{dydt}\\
\frac{dz}{dt} &=& -2\frac{y}{x^2},\label{dzdt}
\eee
Eqs \eqref{dydt}-\eqref{dzdt} result in
\bee
y^2+z^2 &=& c^2.\label{circ}
\eee
Combining Eqs~\eqref{dxdt}-\eqref{dydt} to a single ODE, and solving it, we obtain the relation,
\bee
x&=&\left(1-\frac{y^2}{2c^3}\right)^{-1}.\label{x}
\eee

The time-dependence can be found by solving e.g. the ODE for 
$\lambda=y/z$, 
with $\lambda=0$ at $t=0$,
\bee
\frac{d\lambda}{dt}&=& 2 \left( \frac{c-c_0}{c}\right)^2 \left(\lambda^2+\frac{c}{c-c_0}\right)^2 \frac{1}{1+\lambda^2},\hspace{0.6cm}
\eee
where
\bee
c_0=0.5.\label{ac}
\eee
The solutions are,
\bee \hspace{-0.2cm}
t\!\!&=&\!\! \frac{(1\!-\!a^2)a^2\lambda}{4(\lambda^2\!+\!a^2)}+\frac{a(1\!+\!a^2)}{4}\arctan\frac{\lambda}{a},\;\;\;\;\;\;\mbox{ for } c>c_0,\nonumber\\\label{b1}\\
t\!\!&=&\!\!\frac{\lambda}{2}\left(1+\frac{\lambda^2}{3}\right)\hspace{3cm}\mbox{ for } c=c_0,\\
t\!\!&=&\!\! 
\frac{(1\!+\!a^2)a^2\lambda}{4(-\lambda^2\!+\!a^2)}+\frac{a(1\!-\!a^2)}{4}\text{arctanh}\frac{\lambda}{a},\;\;\;\;\;
\mbox{ for } c<c_0,\nonumber\\
\eee
where  
$|\lambda/a|<1$ and
\bee
a=\sqrt{\left|\frac{c}{c-c_0}\right|}.\label{bl}
\eee

Periodic solutions exist only for $c>c_0$. Otherwise, the horizontal positions (i.e. $\lambda \rightarrow \infty$), are not reached. For $c=c_0$, they would correspond to $t\rightarrow \infty$, and for $c<c_0$, the limit of $t\rightarrow \infty$ results in a finite positive value $\lambda =a$. For $c< c_0$, the system separates into two groups, in which the particle line-of-center with time approaches the inclination determined by $\lambda \sqrt{(c_0-c)/c}$. 

Therefore, two different types of the dynamics exist, and they are separated from each other by a critical initial aspect ratio
$c_0=0.5$.

For $c>c_0$, the limit $\lambda \rightarrow \infty$ takes place when $t \rightarrow T/4$, with the period of the oscillations,
\bee
T &=& \frac{\pi}{2}\sqrt{\frac{c}{c-c_0}}\left(\frac{c}{c-c_0}+1 \right).\label{aT}
\eee
The maximal width of the trajectory, $x=d_{\text{max}}$ is observed at $T/4$ when $y=c$. Therefore, from Eq.~\eqref{x} we obtain,
\bee
d_{\text{max}}&=&\frac{c}{c-c_0}+c.\label{ad}
\eee

The power law divergence of the approximate dynamics in the limit of $c\rightarrow c_0$, seen in Eqs~\eqref{aT} and \eqref{ad}, is in agreement with the analogical scalings \eqref{Tc-c0}-\eqref{ww}, observed for the original equations of motion \eqref{dy}. 

For $2N\!\!=\!\!4$, the approximate value of $c_0$, given by Eq.~\eqref{ac}, is underestimated by around ten percent. This is reasonable taking into account that the exact value is only slightly smaller than unity, and therefore one cannot expect the adopted approximation to be very precise. We are now going to demonstrate that the higher $N$, the better is the accuracy. 


Generalization of the approximate dynamics for a larger number of particles $2N$ is straightforward (for simplicity, we assume that $N$ is even). For the positions of the twin particles $1$ and $N+1$ we now keep the same symbols as in Eq.~\eqref{no} for the particles $1$ and $3$ (in case of $2N=4$), but with the primes added,
\bee
x'=x'_1+x'_{N\!+\!1},\hspace{0.5cm}y'=x'_1-x'_{N\!+\!1},\hspace{0.5cm}z'=z'_{N\!+\!1}-z'_1,\nonumber\\
\label{no'}
\eee
and the particle positions are denoted as ${\bf r}'_i=(x'_i,0,z'_i)$.

The initial conditions are given by Eq.~\eqref{in} in the adjusted notion, i.e. ${\bf r}'_i$ rather than ${\bf r}_i$ and $c'$ rather than~$c$. 
We start from flat configurations with the aspect ratio much smaller than the size of the polygon side, 
\bee
c'<<\sin\frac{\pi}{N},
\eee 
and assume that the twin particles team up and all time stay separated from the other neighbors, 
\bee
y',\;z' << x'\sin\frac{\pi}{N},\hspace{0.8cm}
\eee
The resulting approximate dynamics of the relative coordinates has the form, 
\bee
\frac{dx'}{dt'} &=& 2\frac{y'z'}{c'^3},\label{dxdt'}\\
\frac{dy'}{dt'} &=& 2\frac{z'}{x'^2}(1+\alpha),\label{dydt'}\\
\frac{dz'}{dt'} &=& -2\frac{y'}{x'^2}(1+\alpha),\label{dzdt'}
\eee
where
\bee
\alpha&=& 2\sqrt{2}\sum_{k=2}^{N/2}\frac{1}{\sqrt{1-2x'_{k}(0)}},
\eee
and
\bee
x'_{k}(0)=\frac{1}{2}\cos\frac{2\pi(k-1)}{N}.
\eee
From Eqs \eqref{dydt'}-\eqref{dzdt'} it follows that the distance between the twin particles is constant,
\bee
y'^2+z'^2 = c'^2.\label{circ}
\eee

We now substitute,
\bee
\!\!\! x'=x,\hspace{0.45cm}y'=\frac{y}{1\!+\!\alpha},
\hspace{0.45cm}z'=\frac{z}{1\!+\!\alpha},\hspace{0.45cm}t'=\frac{t}{1\!+\!\alpha},\hspace{0.5cm}\label{sc}
\eee
and recover for $x,y,z,t$ the same dynamics as in Eqs~\eqref{dxdt}-\eqref{dzdt}, but with the  parameter $c$ rescaled as, 
\bee
c'=\frac{c}{1\!+\!\alpha}.\label{c'}
\eee
The solution immediately follows from the benchmark  Eqs.~\eqref{x}, \eqref{ac}
-\eqref{bl} and the scalings \eqref{sc}-\eqref{c'}. 
In particular, it is easy to find the general expression for the critical aspect ratio, which separates two different types of the dynamics: with and without periodic oscillations, 
\bee
c'_0=\frac{1}{2(1\!+\!\alpha)}.
\eee
For example, we consider $2N$=16. In this case,
\bee
\alpha\!\!&=&\!\!2\sqrt{2}\left[1+2^{1/4}\left(\sqrt{\sqrt{2}+1}+\sqrt{\sqrt{2}-1}\right) \right],\hspace{0.8cm}
\eee
and
\bee
c'_0\!\!&\approx&\!\! 0.0446,
\eee
in a very good agreement with the numerical value given in Eq.~\eqref{c0_16}.

\section{Spherical particles}\label{A}
The results presented in the previous sections have been obtained within the point-particle model. In this section, we will show that this approximation well reproduces generic features of the dynamics of spherical particles, even in case when $d$ is only slightly smaller than $c$. 

Consider now $K$ identical spheres moving in a viscous fluid under gravitational 
forces $\bF$. The fluid velocity ${\bf v}$ and pressure 
$p$ satisfy the Stokes equations 
with the stick boundary conditions at the sphere surfaces.
Dynamics of the translational motion of the spheres reads 
\bee
{\dot{\br}}_i(t)  &=& \left[ \sum_{k=1}^K\m_{ik}\right] \cdot  \bF, 
\hspace{1cm} i=1,...,N,\label{dyn}
\eee
where $\br_i(t)$ are time-dependent positions of the sphere centers and 
the mobility matrices $\m_{ik}$ (which depend on 
relative positions of 
all the particles) are evaluated 
numerically by the multipole expansion~\cite{CFHWB,MEJ-EW} with the use of the {\sc Hydromultipole} numerical code~\cite{CE-JW}.

To illustrate periodic gravitational settling of $2N$ identical spheres, we consider the initial configuration of their centers given by Eq.~\eqref{in} with $2N\!=\!16$ and $c\!=\!0.9$. A rather small diameter $d=0.19$ is chosen to keep in balance hydrodynamic interactions between the twin particles and between the closest neighbors from the polygon. 

\begin{figure}[h]
\psfrag{CM,1-16}{}
\psfrag{y}{$x_{i,CM,1-16}$}
\psfrag{z}{$z_{i,CM,1-16}$}
\includegraphics[width=7.5cm]{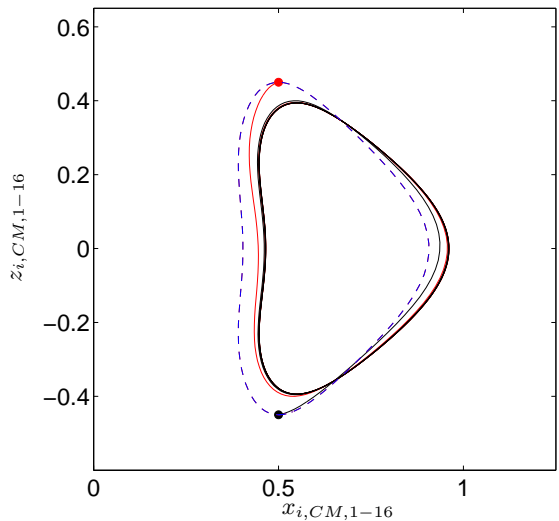} 
\vspace{-0.3cm}
\caption{
Trajectories of centers of two twin spheres from the regular group. Dashed line: without a tail. Solid line: in the presence of a tail-sphere, initially centered at the distance 1.25 above the group center-of-mass. The sphere diameter $d=0.19$. The group center-of-mass frame is taken.
}
\label{zogonem}
\end{figure}

In the center-of-mass frame, the trajectory of the spheres (dashed line in Fig.~\ref{zogonem}) is very close to the point-particle trajectory (solid line in Fig.~\ref{per1}). 
Similarly as in Sec.~\ref{IV}, we now investigate 
how does the shape of the relative trajectory change in time in the presence of an additional identical sphere (``tail''), 
initially located above the group at its symmetry axis at (0,0,1.25), see the solid lines in Fig.~\ref{zogonem}. The initial positions of the sphere centers are  indicated by dots.

Time evolution of the solid lines in Fig~\ref{zogonem} shows that,  owing to the hydrodynamic interaction with the tail, the width of the configuration has become larger, the height smaller, and the hole radius has increased, both in comparison to the initial condition and to the periodic trajectory without the tail (dashed line), in agreement with our findings in Sec.~\ref{IV} for the analogical point-particle system.

\end{document}